\documentstyle[12pt]{article}
\makeatletter
\begin{document}
\topmargin 0pt
\oddsidemargin -3.5mm
\headheight 0pt
\topskip 0mm \addtolength{\baselineskip}{0.20\baselineskip}
\begin{flushright}
 hep-th/9910213
\end{flushright}
\vspace{0.5cm}
\begin{center}
  {\large \bf Comments on `` Entropy of 2D Black Holes from Counting 
Microstates '' }

\end{center}
\vspace{0.5cm}
\begin{center}
by 
\vspace{0.3cm} \\
  Mu-In Park\footnote{Electronic address:
    mipark@physics.sogang.ac.kr} and Jae Hyung Yee
\\ 
  {\it Department of Physics, Yonsei University, Seoul 120-749, Korea } \\ 
\end{center}
\vspace{0.5cm}                              
\begin{center}
  {\bf ABSTRACT}
\end{center}
We point out that a recent analysis by Cadoni and Mignemi on the statistical 
entropy of 2D black holes has a serious error in identifying the Virasoro algebra 
which invalidates its principal claims.

\vspace{10cm}
\vspace{0.5cm} 
\begin{flushleft}
PACS Nos: 04.70.Dy, 11.25.H, 11.30.-j, 04.20.Fy\\
9 September 1999 \\
\end{flushleft}
\newpage

Recently Cadoni and Mignemi \cite{Cad:99} presented a microscopical derivation 
of the entropy of the two-dimensional (2D) black holes. They have shown that the 
canonical algebra of the 
asymptotic symmetry of 2D Anti-de Sitter (ADS) space included the Virasoro 
algebra. Using this
algebra and the Cardy's formula they obtained the statistical entropy which agreed, up 
to a factor of $\sqrt{2}$, with the thermodynamical result. In this Comment, we point out 
that their analysis has a serious error which invalidates its principal claims.  

Let us consider the problem in detail. The authors' object was obtaining the 
microscopic derivation of the entropy of the 2D black holes in ADS 
space \cite{Cad:95} from the microstate counting procedure of the 
three-dimensional ADS space $a~`la$ 
Brown-Henneaux \cite{Bro:86} and Strominger \cite{Str:98} (BHS). In this approach, it 
is a basic ingredient that there is the infinite-dimensional conformal symmetry,
for some (infinitely large or finite) boundaries, which is described by the 
Virasoro algebra with the ``classical'' central charge \cite{Bro:86,Ban:96}. It 
is the strategy to apply the Cardy's formula for the asymptotic density of states 
to obtain the microscopic entropy though the true validity of the formula may 
be debatable \cite{Car:98}.
  
As the simplest example of the 2D gravity theory that admits 2D black hole solution in ADS 
space they considered the Jackiw-Teitelboim model \cite{Jac,Cad:95}. Using 
the Regge-Teitelboim procedure \cite{Reg:74}, they obtained the differentiable, 
which means `no boundary terms in the field variation', Hamiltonian
$$
H[\chi]=\int dx (\chi^{\perp } {\cal H} +\chi^{\parallel} {\cal H}_x ) + J[\chi]
$$
for the surface deformation parameters $\chi^{\perp}=N \chi^t, 
\chi^{\parallel}=\chi^x + N^x \chi^t$ with the lapse, shift functions $N, N^x$ 
and the Killing vectors $\chi^t, \chi^x$ along the tangent vectors 
${\partial}/{\partial t},{\partial}/{\partial x}$ respectively. 
Here, note that the boundary charge $J[\chi]$ is defined only at a boundary 
``point'', $x \rightarrow \infty$, and no other integration variables present. With this Hamiltonian, 
the central charge are read off usually from the Dirac bracket algebra 
\cite{Bro:86} 
\begin{eqnarray}
\{J[\chi], J[\omega]\} _{DB} =J[[\chi, \omega]]+c(\chi, \omega)
\end{eqnarray}
or from the variation of $J[\chi]$ under the surface deformations 
\begin{eqnarray}
\delta_{\omega} J[\chi]=J[[\chi, \omega]]+ c(\chi, \omega)
\end{eqnarray} 
with the surface deformation Lie bracket $[\chi,\omega]$ \cite{Bro:86} and 
its corresponding central term $c(\chi, \omega)$. [The Dirac bracket for the 
dynamical variables $A, B$ is defined usually as $\{A, B \}_{DB}=\{A, B\} - 
\{A, \Gamma_{\alpha}\} C^{-1}_{\alpha \beta} \{\Gamma_{\beta}, B \}$ with the 
second-class constraints $\Gamma_{\alpha} (\mbox{det} C_{\alpha \beta} \neq 0, C_{\alpha \beta}\equiv 
\{\Gamma_{\alpha}, \Gamma_{\beta}\})$ and 
the inverse of $C_{\alpha \beta}$ , $C^{-1}_{\alpha \beta}$ \cite{Dir:64}. 
In our case of (1), 
it is {\it implicitly} assumed that the gauge-fixing conditions 
$F_{\gamma}\approx 0$ are introduced to make the 
(first-class) energy and momentum constraints ${\cal H} \approx 0, {\cal H}_x \approx 0$ become the second-class 
constraints set $\Gamma=\{{\cal H}, {\cal H}_x, F_{\gamma}\}$.]  

But there are serious mismatches \cite{Park:99}
with our purpose: a) Because the algebra is defined only at one (boundary) point 
there is no room for the infinite tower of symmetry generators through the 
Fourier-series expansion of the algebra as required in the BHS procedure. b) 
$J[\chi]$ is time-dependent in general which means that $J[\chi]$ is not 
conserved quantity. This is contrast to the usual fact that $J[\chi]$ is a 
(boundary part of) conserved Noether charge \cite{Ban:96}. Moreover, the algebra 
and the central term $c(\chi, \omega)$ are now time-dependent and eventually 
``time-dependent entropy'' would be expected $S \sim \sqrt{c(\chi, \omega)}$ 
according to the Cardy's formula \cite{Cad:99,Car:98} in general. In order to resolve this 
unwanted situation they introduced the time-integrated charge $\hat{J}$ 
\begin{eqnarray}
\hat{J}[\chi] =\frac{\lambda}{2 \pi} \int^{2 \pi/\lambda}_0 dt J [\chi]
\end{eqnarray}
with time period of $2 \pi/\lambda$ and they obtained a Virasoro algebra in the 
asymptotic symmetry algebra of 2D black holes. But this procedure is an 
erroneous one. Let us explain our argument in detail.
  
We start our argument by considering Eq. (2) in their suggesting frame. It is true that
the (overall) time integration of (2) has a definite meaning as 
\begin{eqnarray}
\widehat{~~\delta _{\omega} J[\chi]~~ } = \hat{J}[[\chi,\omega]] 
+\hat{c}(\chi, \omega)
\end{eqnarray}
with the central charge (23) of Ref. [1] and this looks like a Virasoro algebra. But 
the problem of this method is that the left-hand side of 
(4) can not be written as
\begin{eqnarray}
\{\hat{J} [\chi],\hat{J} [\omega] \}_{DB},
\end{eqnarray}
which is essential for the interpretation of the time integrated (1) as a Virasoro algebra: Let us consider 
\begin{eqnarray}
\{ J[\chi], H[\omega] \}_{DB} =\{J[\chi], J[\omega]\}_{DB} = \delta_{\omega} J[\chi]
\end{eqnarray}
which implies that $H[\omega]$ is the correct (asymptotic) symmetry generator, i.e., 
\begin{eqnarray}
\{\phi(x), H[\omega]\}_{DB}=\delta_{\omega} \phi (x).
\end{eqnarray} 
for the symmetry transformation $\delta_{\omega} \phi$ of 
the field $\phi$. [This Dirac bracket as well as the Poisson bracket in Eq. (19) of Ref. \cite{Cad:99} can 
be well-defined contrast to the claims of Cadoni and Mignemi following the work 
of Ref. 
\cite{Park:98}.] Here it is an important fact that the Dirac brackets are all defined at 
$equal~times$ because the constituting Poisson brackets are equal times in 
nature for the $local$ field theory: 
$$
\{A, B\}=\int dx \left(\frac{\delta A}{\delta \phi_k (x,t)}\frac{\delta B}{\delta \pi^k(x, t)} 
-\frac{\delta A}{\delta \pi^k(x,t)}\frac{\delta B}{\delta \phi_k(x, t)} \right) 
$$
for the canonical conjugates pairs $\phi_k(x, t), \pi^k (x,t)$.
If we perform the (overall) time integration over (3) the 
left-hand side becomes
\begin{eqnarray*}
\frac{\lambda}{2 \pi} \int^{2 \pi/\lambda}_{0} dt \{ J[\chi(t)], J[\omega(t) ]\}_{DB}
\end{eqnarray*}
or
\begin{eqnarray*}
\left(\frac{\lambda}{2 \pi}\right)^2 \int^{2 \pi/\lambda}_{0} dt'  
\int^{2 \pi/\lambda}_{0} dt \frac{2 \pi}{\lambda} \delta(t-t') 
\{ J[\chi(t)], J[\omega(t) ]\}_{DB}.
\end{eqnarray*}
It is then clear that (6) can not be written as (5) because of the factor 
$\frac{2 \pi}{\lambda} \delta(t-t')$ in the integrand. This is a direct consequence of 
(6) which is a well-known fact related to the Noether theorem in the field theory; if 
their claim which equating (4) and (5) was correct, the time-integrated quantity 
$\hat{H} [\chi]~(\approx \hat{J} [\chi] )$ should be treated as the symmetry generator 
necessarily contrast to (7).
 
In conclusion, they made a serious mistake by identifying (4) and (5) and 
hence their way of applying the Cardy's formula to obtain the black hole 
entropy (25) of Ref. [1] with the misidentified $c$ (central charge) and $l_0$ 
(lowest eigenvalue of the Virasoro generator) can not be justified.

MIP would like to thank Prof. Steven Carlip, Drs. Gung-won Kang and Hyuk-jae 
Lee for several helpful discussions. MIP was supported in part by a 
postdoctoral grant from the Natural Science Research Institute, Yonsei 
University in the year 1999 and in part by the Korea Research Foundation under 
98-015-D00061.

\end{document}